# Two-Parameter Quasi-Ballistic Transport Model for Nanoscale Transistors

Ji Ung Lee[1]*, Ramya Cuduvally[1], Prathamesh Dhakras[1], Phung Nguyen[1], and Harold L. Hughes[2]

[1]Colleges of Nanoscale Science and Engineering, SUNY-Polytechnic Institute, Albany, NY, 12203, USA. [2]US Naval Research Laboratory, Washington, DC, 20375, USA.

***Abstract*— We show that by adding only two fitting parameters to a purely ballistic transport model, we can accurately characterize the current-voltage characteristics of nanoscale MOSFETs. The model is an extension of Natori's model (J. Appl. Phys. 76, 4879 (1994)) and includes transmission probability and drain-channel coupling parameter. The latter parameter gives rise to a theoretical $R_{ON}$ that is significantly larger than those predicted previously. To validate our model, we fabricated n-channel MOSFETs with varying channel lengths. We show the length dependence of these parameters to support a quasi-ballistic description of our devices.**

## I. INTRODUCTION

In nanoscale conductors that have channel lengths comparable to the mean free path of carriers, transport approaches the ballistic limit. A ballistic model should also describe the transport properties of well-designed MOSFETs with channel lengths approaching or below the mean scattering length λ. On the other hand, there is a marked difference between the current-voltage ($I_D$-$V_D$) characteristics from a purely ballistic model, such as the one proposed by Natori[1], and experimental data. Although complex models can fit experimental data more precisely[2], the simplicity of a fewer-parameter model based on a ballistic transport formalism is desirable.

In Natori's model for ballistic transistors, MOSFET current-voltage relations were derived based on Landauer's formula for current through a ballistic conductor. In this work, we extend this transport model by including two physically meaningful parameters to accurately predict device data. They are transmission probability $T$ and drain-channel coupling parameter $\Delta$. $\Delta$ captures the effects of the longitudinal electric field in the channel and represents the fraction of the drain Fermi level that appears on the source-end. The inclusion of $\Delta$ allows us to match the experimental ON-resistance ($R_{ON}$) in the low bias regime, which is the central result of our work. We provide physical justifications for these parameters, which hitherto had not been incorporated in a ballistic model to our knowledge.

Additionally, the threshold voltage ($V_T$) is either obtained from device characterization techniques or allowed to vary so as to account for $V_T$ variation with device length. Here, we use either a two-parameter fit, by fixing $V_T$, or a three-parameter fit, where all three parameters are allowed to vary, to characterize experimental data. By allowing $V_T$ to vary, we can account for the effects of the halo implants on the channel length, for example. We relax this requirement when comparing devices over a narrower range of length where the $V_T$ variation is small.

To support our model, we fabricated n-channel MOSFETs with gate lengths $L$ ranging between 50nm to 3μm. We provide a detailed length dependence of the parameters we examined, including $T$ and $\Delta$. Using these parameters, we predict the minimum ON-resistance ($R_{ON}$) that can be achieved theoretically by setting $T$=1. This value is substantially higher than that predicted from a purely ballistic model from others. We also model our data using the simplest long-channel model. By comparing the two models, we calculate the demarcation length at which one model becomes more appropriate over the other. Most importantly, we find that $T$ can be assumed to be independent of drain bias. A bias independent $T$ is counter to the conclusions from the virtual source (VS) model, which has been successful in modeling nanoscale MOSFETs [2-5]. Here, we provide justifications for using a fixed $T$ and discuss the assumptions used in each model that determine their respective values for $T$.

## II. Ballistic Transport

### A. Landauer's formula

In nanoscale transistors, where the channel length becomes comparable to or less than the mean free path of carriers, we enter the quasi-ballistic or pure ballistic regime of transport. Here, parameters like the mobility and mean free path start to lose the physical significance that they held in the diffusive transport regime. In a ballistic transistor, scattering events are ideally zero. Landauer emphasized the role of contacts in determining the current through such a conductor. The maximum conductance, in this case, is limited by the contacts and is given by $G_c = 2q^2/h$ per conduction mode, where $q$ is elementary charge and $h$ is Planck's constant. Thus, the resistance is finite even though there is no scattering in the conductor. The minimum resistance arises because the current is carried by many transverse modes in the contacts but by only a few modes or subbands inside the conductor. This requires a redistribution of the current among the current carrying modes at the interface that leads to a finite contact resistance.[6] It can also be thought of as an energy penalty that an electron has to pay in order to be funneled into the narrow channel from contacts that are comparatively wide. Here, we distinguish between this contact resistance, an intrinsic parameter that sets the ultimate resistance of a ballistic device, and series resistance, an extrinsic effect that arises from the physical size and nature of the contact to the source/drain regions of a MOSFET.

In this report, we describe a new ballistic transistor model with the fewest fitting parameters reported today. The important insight we provide is that the net occupancy of transmission modes is reduced from the original Natori's model due to a finite longitudinal electric field in the channel, which is captured in the parameter $\Delta$. In the Appendix, we describe the origin $\Delta$ in greater detail.

### B. Natori's Model for Ballistic Transistors

In Landauer's formalism (Eq. 1) for ballistic transport[7], the current through any conductor is proportional to the probability $T(E)$ that an electron can transmit through it and the quasi-Fermi levels that determine the occupancy of right- and left-moving carriers. In a conductor with multiple subbands, the current also depends on the number of modes *M(E),* which depends on energy $E$. Hence, the difference in the Fermi functions $f_1 - f_2$ determines the net transmission modes.

$$I_D = \frac{2q}{h}\int dE\, T(E)M(E)(f_1 - f_2) \tag{1}$$

Using Landauer's approach, K. Natori[1] derived the current-voltage ($I_D$-$V_D$) characteristics of a ballistic MOSFET. Natori's model was developed for purely ballistic transistors that have channel lengths $L< \lambda$, where $\lambda$ is the mean scattering length, and for perfect gate coupling. Even by setting T<1, however, the model does not represent experimental data. Fig. 1 shows Natori's model with T<1 to match the maximum current in our $L$= 60nm device, which we show again in Fig. 3(b) with our modified model. In Fig. 1, it is clear that the ON-resistance ($R_{ON}$) in the low bias regime is unrealistically too small compared to the experimental data.

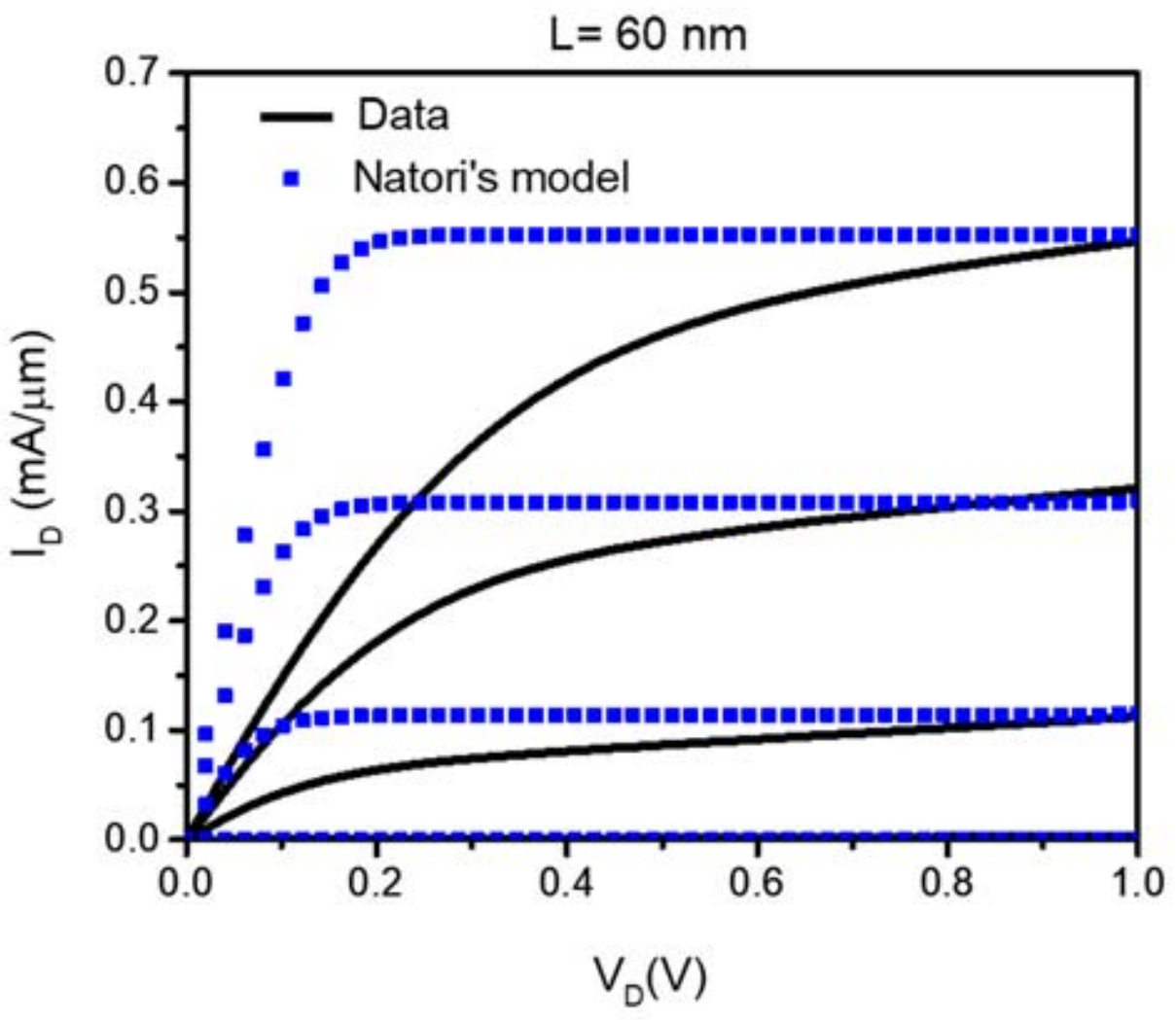

**Fig. 1. Natori's model, with T<1 used to estimate device data, predicts an unrealistically low $R_{ON}$ at low drain bias. The same device data is modeled in Fig. 2 using the quasi-ballistic model described in the text.**

The original model, however, serves as a benchmark for the highest current that could be achieved for a given technology node. For example, the degree of ballisticity that one can extract from it is an extremely useful parameter for assessing the scaling potential of transistors. Here, by extending the original model, we are able to better match to experimental data and provide new insights into the nature of ballistic transport in highly scaled MOSFETs.

Eq. 1 clearly shows the competition between $T(E)$ and $M(E)(f_1 - f_2)$. To determine them, one must first realize that these parameter are local, as discussed in more detail in the Appendix. Therefore, although Eq.(1) can be determined at any point along the channel, we apply it at a point that allows us to gain the most physical insight. This point is at the top of the band near the source, as we discuss below. At the top of the source, we must carefully define the quasi-Fermi levels which may be different from those of the source/drain contacts, as we discuss in the Appendix in more detail. When these parameters are properly accounted for, we show that a simple two-parameter quasi-ballistic model for nanoscale MOSFETs can be developed that match experimental data. Our results show that $T(E)$ need not depend on bias, unlike the previous results[5,8-12]. We show that this discrepancy fundamentally arises from the use of non-local parameters to determine the net occupancy of $M(E)$.

## III. A TWO-PARAMETER BALLISTIC MODEL

In this work, we present a simple yet physical modifications to the ballistic transistor model to match experimental data. This is accomplished by extending Natori's ballistic model to include the effects of channel transmission ($T$) and drain-channel coupling parameter $\Delta$. These modifications are supported by the physics of the device that we present. In this section, we discuss the origin of these two parameters in detail. We show that $T$ and $\Delta$ are closely linked. Depending on the use of either local or nonlocal quasi-Fermi levels, a bias independent or a bias dependent $T$ will emerge.

### *A. Transmission (T)*

In a quasi-ballistic MOSFET, the channel length is comparable to the mean free path λ of carriers in the channel. Thus, carrier scattering is significantly reduced but is not entirely absent. Most importantly, mobility no longer becomes a valid parameter. Instead, transmission ($T$) captures the effect of these scattering processes on the drain current. The most natural place to assess the value for $T$ is at the peak of the band diagram[8], as we show in Fig. 2. This is because the carrier density there is controlled by the gate and the lateral field is small. This region is close to the source and in near equilibrium with the source quasi-Fermi level. Thus the source quasi-Fermi level determines the occupancy of the right-moving carriers. Since the net current is due to the difference in the right- and left-moving carriers, one must also determine the occupancy of the left-moving carriers, which we discuss below.

In our model, $T$ is largely independent of drain voltage and energy, although this assumption is not always found in the literature [5,9,10]. This discrepancy in $T$ arises because it enters in the Landauer's model as a product with $M(E)$. Our model captures the imperfect coupling of the gate to the channel. In such cases, a longitudinal field develops due to the drain bias, and the drain quasi-Fermi level necessarily has to vary along the channel, as discussed in the Appendix. In other words, $(f_1 - f_2)$ must be defined locally, not from those of the source/drain contacts as is done typically. For example, in the model of ref. [5], the drain quasi-Fermi level is assumed to be constant along the channel and was used to model the current at the top of the source. This is the same assumption used in Natori's original work. When the drain quasi-Fermi level is used throughout the channel, however, one underestimates the occupancy of the left-moving modes at the source-end. Since the total current is the difference in the right- and left-moving carriers, the overall current is larger than if one allowed the drain quasi-Fermi level to vary along the channel. As stated earlier, a varying quasi-Fermi level is necessary due to the longitudinal field and requires one to define the quasi-Fermi levels locally. When the drain quasi-Fermi level is used throughout the channel, however, a bias dependent $T$ is needed to match experimental data to compensate for a reduced number of left-moving modes[5]. Typically, $T$ has to start out small and increase with the drain bias $V_D$. In essence, since the product $T(E)M(E)$ enters into the Landauer model, we have shifted the bias dependence to the occupancy of $M(E)$ while previous models have incorporated the bias dependence in the parameter *T*. Clearly, the two approaches result in a completely different physics of nanoscale MOSFETs, even though both may reproduce experimental data.

Here, we show that device data can be modeled accurately using a model that does not require *T* to vary with $V_D$. In the simplest form, *T* is given by Eq. (2).

$$T = \frac{\lambda}{\lambda + L} \qquad (2)$$

$$I_D/W = TI_0(F_{1/2}(u) - F_{1/2}(u - v_d)) \qquad (3)$$

Eq (3) is the $I_D$-$V_D$ characteristics given by Natori (Eq. 22 in ref. 1), modified by *T* and $v_d$. (We note that $v_d$ needs to be normalized to the thermal voltage before using it in Eq. (3)). The first term in (3) captures the current in the channel due to right-moving carriers and the second term captures the current due to left-moving carriers. *W* is the width of the device and $I_0$ is determined by material parameters. $F_{1/2}$ is the half order Fermi integral that can be evaluated numerically. The difference in the two $F_{1/2}$ functions accounts for the nonequilibrium carrier density in the channel under bias. As stated earlier, the most natural place to assess the transport is at the top of the conduction band, as shown in Fig. 2.[1,4,13] Here $v_d$, instead of the actual drain bias $V_D$ , is used to determine the occupancy of the left-moving carriers at the top of the source. We define $v_d$ empirically as a fraction $\Delta$ of the drain bias $V_D$.

*T* is a fitting parameter that we extract for different length devices. Due to scattering, the ON current of a MOSFET is less than the ballistic ON-current by the fraction $T < 1$ that depends on the relative values of the channel length *L* and λ. We assume that the mean free path λ is a constant for a given transistor node. In addition, the ON-current is also reduced in the linear region because we have a larger number of left-moving carriers due to the use of $v_d$ instead of the drain voltage $V_D$.

When $\Delta$ =1, the Fermi level of the drain contact determines the difference $f_1 - f_2$ in Eq. (1), which is valid only in the case of perfect gate coupling. This is the origin of the unrealistically small $R_{ON}$ predicted in the original Natori's model, as we show in Fig. 1 and discussed elsewhere[14]. We show below that by including $\Delta$ <1, we can achieve a more realistic $R_{ON}$ values.

### B. *Effective Drain-Quasi Fermi Level (Δ)*

In our model, we use $\Delta$ as a fitting parameter to account for the longitudial electric field in the channel. This field modifies the quasi-Fermi level of the left-moving carriers at the top of the source. Since the electrical field there is small, the right moving carriers can be assumed to be in near equilibrium with the source contact. This allows us to use the source quasi-Fermi level $E_{fs}$ to determine the occupancy by the right-moving carriers.

The population of the left-moving carriers at the peak, however, can't be determined by $E_{fd}$ of the drain contact. To elucidate the dynamics at the peak, we superimpose the dispersion relation (*E vs k*) and the conduction band profile under bias in Fig. 2. The difference in the quasi-Fermi levels $E_{fs}$ and $E_{fd}$ due to a bias on the drain $V_D$ creates an imbalance in the populations of right- and left-moving electrons throughout the channel. In a purely ballistic channel with no longitudinal electric field in the channel, $E_{fd}$ would be used to demarcate the quasi-Fermi level of the left-moving electrons throughout the channel, defined as $E_{fd} = E_{fS} - qV_D$. In most MOSFETs where the gate control is not perfect, a potential energy profile takes on the form depicted in Fig. 2(a). To account for the longitudinal field, we define $E_{fd}$' that results in a larger occupancy of the left-moving modes at the top of the source. Here, $E_{fd}$' must be above $E_{fd}$, as depicted in Fig. 2, and determines the population of the left-moving carriers at the top of the band. Even if $E_{fs}$ is not strictly in equilibrium with the source contact, the band bending necessarily implies that $E_{fd}' > E_{fd}$ at the top of the band[6].

(a)

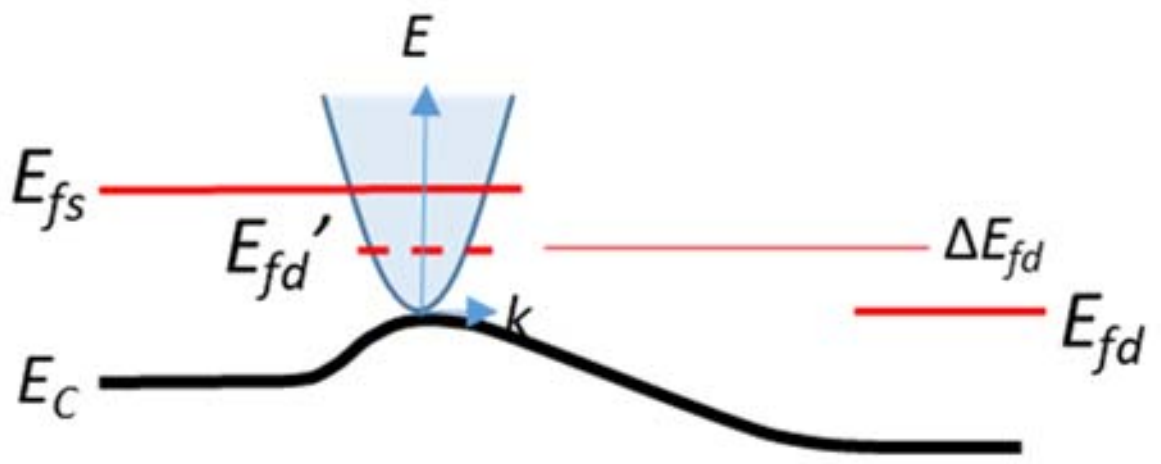

(b)

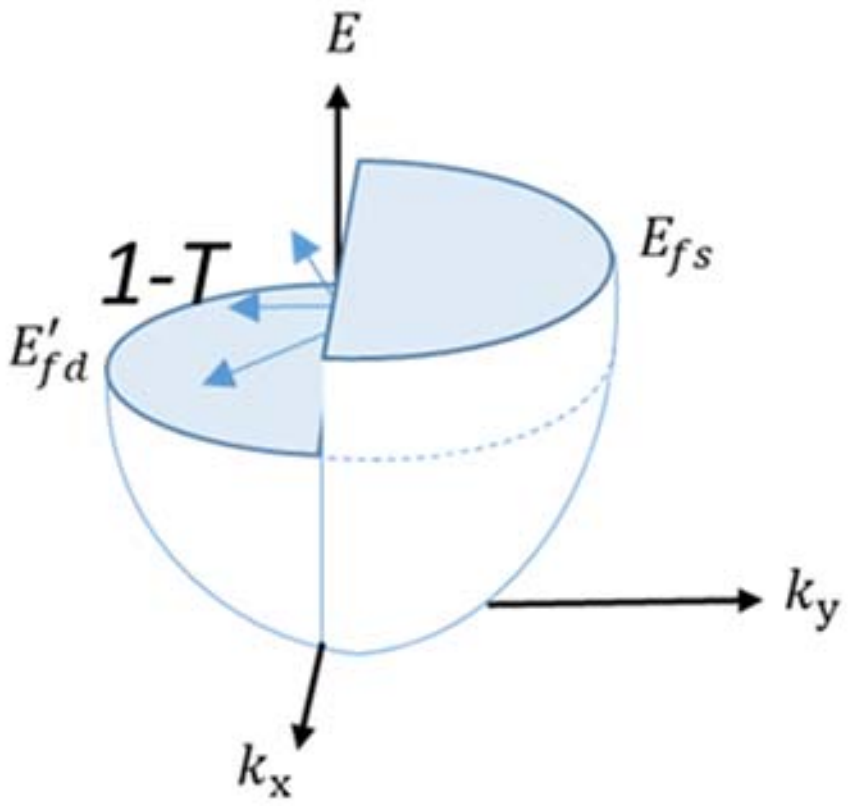

**Fig 2: (a) Band profile along the channel of a quasi-ballistic transistor depicting the effective quasi-Fermi level $E_{fd}$' for the left-moving electrons at the source-end. The factor Δ measures the fraction of $E_{fd}$ and that appears at the top of the source. (b) The full dispersion relation at the top of the band showing that the reflected carriers have equal states as the transmitting carriers to which to reflect, which implies that 1-T is bias independent.**

Just as $E_{fd}$ need to be defined locally, so must $T(E)$. To illucidate, we show the full effective mass dispersion relationship at the peak of the band in Fig. 2(b). Since the lateral field at the top of the source is small, the right- and left-moving carriers are expected to be well defined by local quasi Fermi levels $E_{fs}$ and $E_{fd}$', respectively. As indicated by the arrows, the number of states that are available for the reflected carriers ($1-T(E)$) are equal to the transmitting states for all bias conditions. The preceding argument holds irrespective of whether $E_{fd}$' or $E_{fd}$ is used. Therefore, we can see that as long as the gate keeps the carrier density and the barrier height fixed at the peak, $T(E)$ is expected to be largely independent of $V_D$. In fittng to device data, we show that a single value for $T(E)$ can accurately predict the $I_D$-$V_D$ curves for a given channel length device.

We now turn to estimating $E_{fd}'$ to model experimental data. From Fig. 2, we have elected to use $v_d = \Delta \cdot V_D$ where $\Delta$ is a constant determined by fitting to experimental data. Thus we have $E_{fd}' = -qv_d$. We expect $\Delta$ to vary with transistor geometry; the more control the gate exerts on the channel, the closer $\Delta$ will approach 1. The use of a constant $\Delta$ is furrther justified by fitting to device data. We show that $\Delta$ is a fundamental parameter of our MOSFETs, rather than one that is affected by extrinsic parasitic effects.

In addition to the quasi-ballistic devices, we measured long-channel MOSFETs. Because of the wide variation in the length we examined, a fair amount of threshold voltage variation with channel length is observed across the devices. Thus, initially, $V_T$ is allowed to vary as a fitting parameter. Since this variation is negligible for devices with similar lengths, $V_T$ is later considered to be fixed for those devices. The variation in $V_T$ is due to processing steps incorporated to reduce short-channel effects. For example, with decreasing channel length, $V_T$ roll-off can occur. To counter this, source/drain extensions and halo implants are added. These steps are responsible for the length dependence of $V_T$ with channel length. For devices with $L < 100$nm, we can fix $V_T$ without affecting the fits.

## IV. Results

To support our model, we fabricated n-channel planar MOSFETs at SUNY-Poly's 300mm fabrication facility. The devices have channel lengths that range from $L$=50nm to 3µm, with width $W$=1µm for all devices. The MOSFETs have an equivalent $SiO_2$ gate thickness of 2nm. We determined the fitting parameters using a least-squares minimization.

### A. R-squared Goodness of Fit

In Figs. 3(a) and 3(b), we show the results from $L$=1µm and 60nm MOSFETs. For both devices, we fit the data using a simple long-channel model given in Eq. (4) and the modified ballistic model discussed in section III. We, however, do not expect the long-channel parameters we extract for the $L$=60nm device to be physical. Similarly, we do not expect the parameters from the quasi-ballistic model to be valid for $L$=1µm device. This exercise, however, will help to assess the length scale at which one model is more accurate over the other.

$$I_D/W = \frac{C_{ox}\mu_e}{L}[(V_G - V_T)V_D - V_D^2/2] \qquad (4)$$

For the long-channel Eq. (4), $V_T$ and electron mobility $\mu_e$ are fitting parameters. Beyond the linear region, we assume a constant current given by the saturation current determined from Eq. 4 in the usual way. We treat the two models on an equal footing by initially allowed $V_T$ to vary for both models, although we expect significant discrepancies in the results from the two models and expect some of the parameters to be not physical.

We confirmed a cross-over in the R-squared goodness-of-fit in Fig. 4, which is plotted for the two models as a function of channel length. A cross-over in R-squared values suggests that there is a length scale below which the ballistic description is the more appropriate one

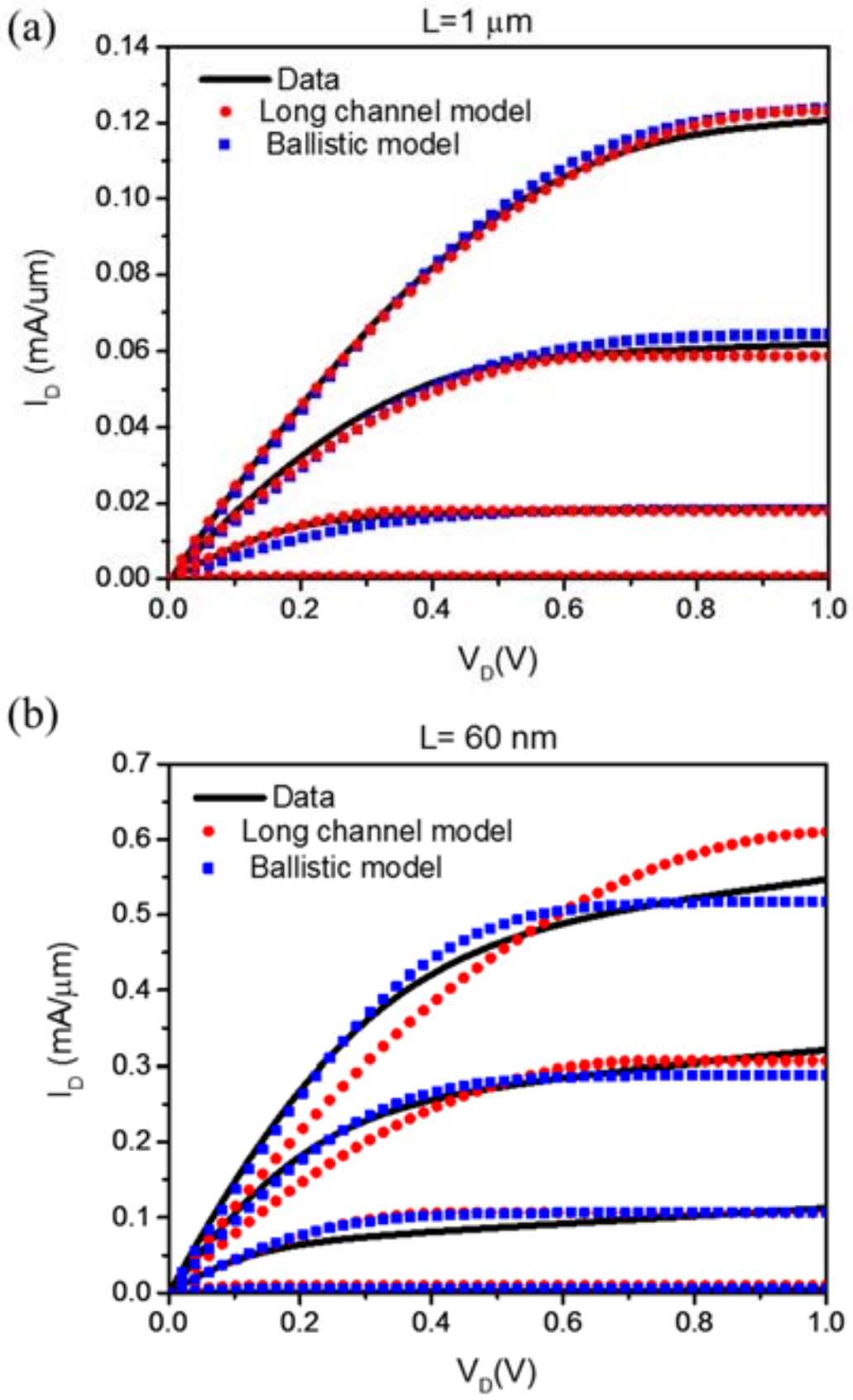

**Fig 3: $I_D$-$V_D$ characteristics of n-channel MOSFETs for (a) $L$=1 μm and (b) $L$=60 nm. The gate voltage is stepped from 0 to 1.2V in 0.3V increment. For each device, we use both a long-channel model and the quasi-ballistic model to fit the data. To treat both models on an equal footing, we allow $V_T$ to vary. (a) From the long-channel model, we find $\mu_e$= 151 cm$^2$/Vs and $V_T$ =0.23 V; from the quasi-ballistic model, we have T=0.05 and $V_T$ =0.44 V. (b) From the long-channel model, we have $\mu_e$ = 46 cm$^2$/Vs $V_T$ =0.19 V; from the quasi-ballistic model, we have T=0.22 and $V_T$ =0.38V.**

For the device with $L$=1μm in Fig. 3(a), both models result in a good fit, provided that $V_T$ is allowed to vary. However, the long-channel model is clearly the correct model for this channel length and results in a better fit as measured from the R-squared goodness-of-fit shown in Fig. 4. The long-channel mobility we extract is $\mu_e$= 151 cm$^2$/Vs. The quasi-ballistic model for this length device gives a reasonably good fit, albeit with a clearly wrong $V_T$. The threshold voltage is over 0.2V above that of the long channel model, which is the correct one as determined from direct measurement (see the figure captions for the extracted parameters).

For the $L$= 60nm device in fig. 3(b), we observe that the quasi-ballistic model provides a far better fit compared to the long-channel model. The long-channel model requires unreasonably low values for both $\mu_e$ and $V_T$, which are not physical [15,16].

In order to determine the channel length below which the quasi-ballistic model would be more appropriate, both the long channel and the quasi-ballistic fits were performed for all channel lengths and the goodness of fit parameter, R-squared, was plotted as a function of length for both fits. This is shown in Fig 4. We see from this figure that a crossover in R-squared occurs around $L$ =500 nm, although this is not the demarcation length for quasi-ballistic transport since we allowed $V_T$ to vary. Indeed, we expect each model to overestimate the range in which the underlying physics is valid because of we allowed $V_T$ to vary. However, there is clear confirmation that our modified quasi-ballistic model is a better model for highly scaled devices below a certain channel length. The devices near the crossover length at 500 nm can be viewed as neither being quasi-ballistic nor purely diffusive.

We thus conclude that the addition of only two parameters to the purely ballistic model from Natori can dramatically improve the $I_D$-$V_D$ characteristics for nanoscale transistors, which we demonstrate for the first time.

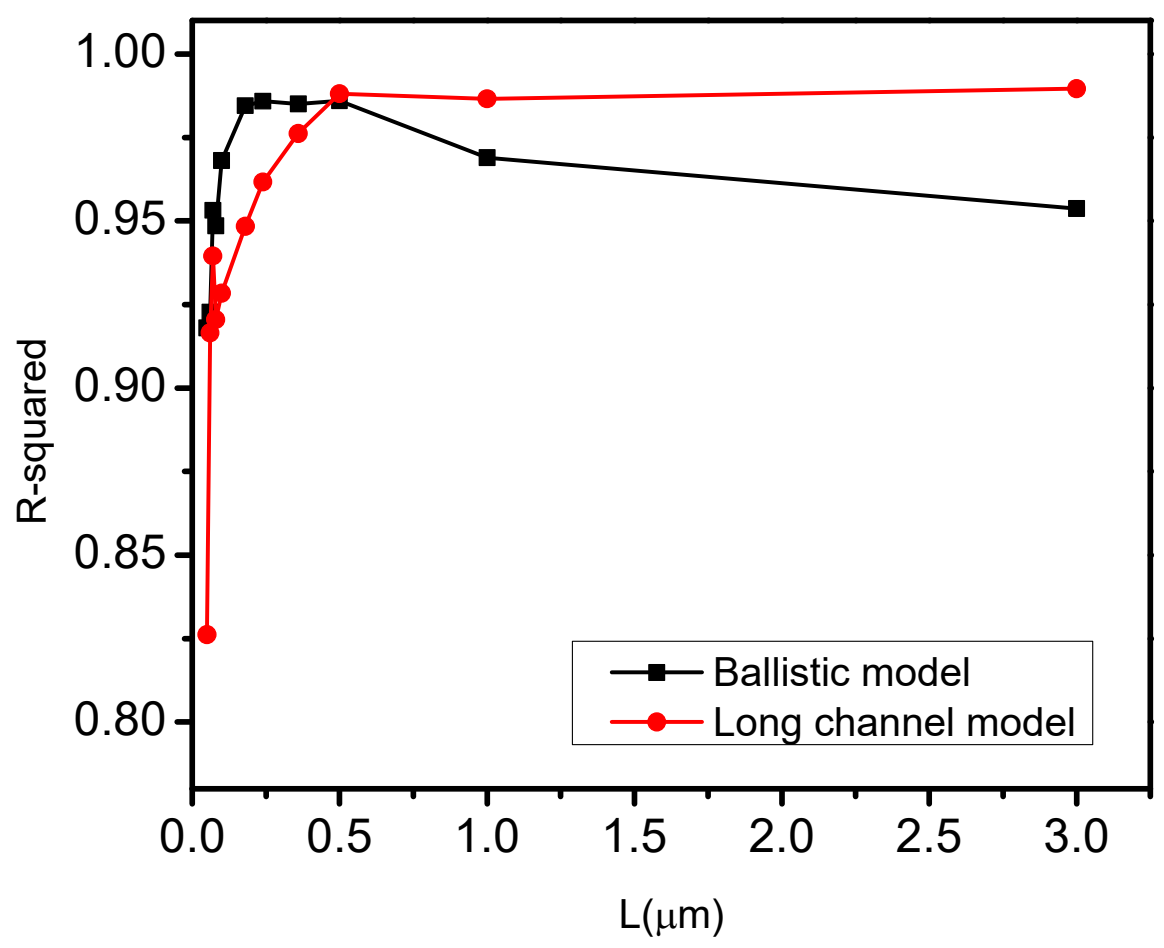

**Fig 4: R-squared goodness of fit vs. length for long channel model (red) and quasi-ballistic model (black) treated equally by allowing $V_T$ to vary. A crossover is seen near *L*=0.5µm.**

*B. Experimental Determination of T and Δ*

To better support the quasi-ballistic model, we examine the length dependence of transmission $T$. In Fig 5, we fix $V_T$ and examine devices over a narrower range of length. From the plot of transmission probability vs. length of all device lengths ranging from 60 nm to 0.5 µm, it was observed that only the devices with $L$ < 100 nm followed the trend given by Eq. (2). To extract the transmission probability, the threshold voltage was fixed to 0.4V, which was determined experimentally for these devices, and a two-parameter fit was performed. Fixing $V_T$ did not affect the quality of the fit for devices with $L$ < 100 nm.

From the fit, we extract λ=19 nm. This implies that the quasi-ballistic model would be more applicable below 100 nm. Therefore, according to the trend in transmission probability with length, a more precise estimate of the demarcation length below which the devices enter the quasi-ballistic regime of transport is around 100 nm. This demarcation length is further supported by other analysis we provide in Figs. 6 and 7.

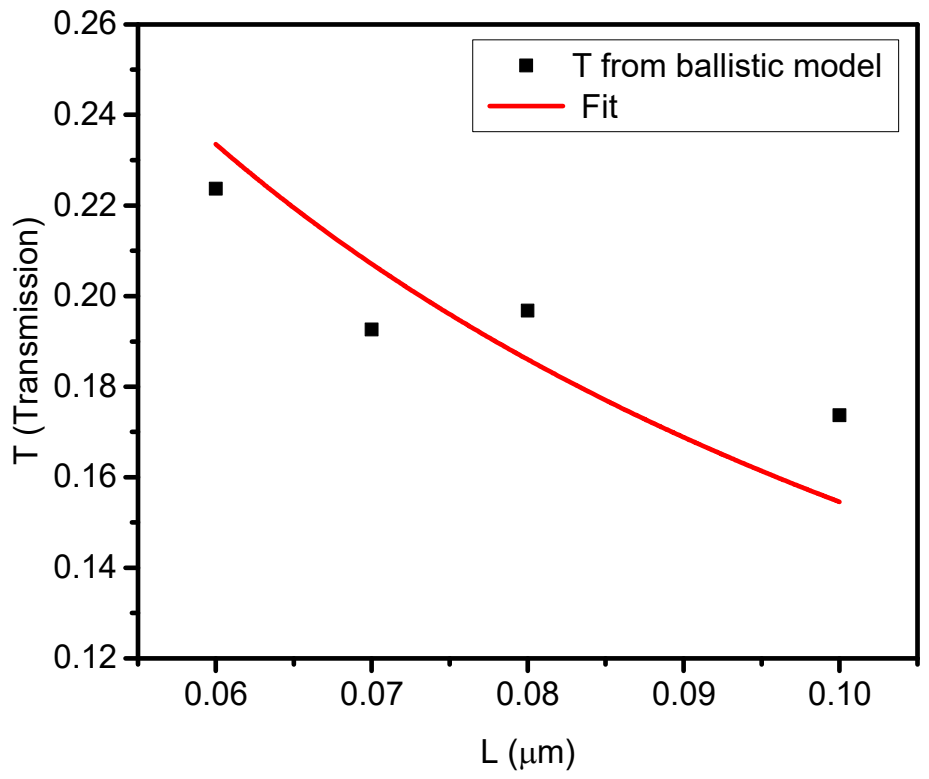

**Fig 5: Transmission probability $T$ vs. Length determined from fitting the $I_D$-$V_D$ curves to the modified Natori's model. The solid curve is a fit to Eq. (2), yielding λ=19 nm.**

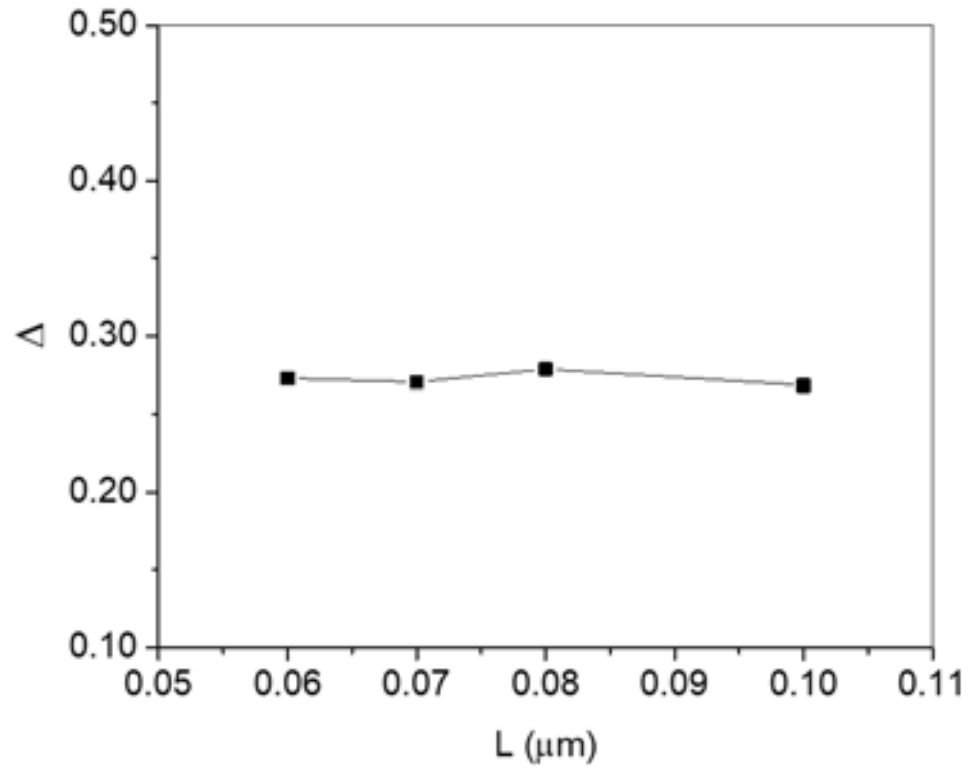

**Fig 6: Extracted $\Delta$ vs length (L) from the fit for the devices examined in Fig. 5.**

We plot the extracted $\Delta$ in Fig. 6 for the devices shown in Fig. 5. The striking result is that $\Delta$ is largely independent of channel length. Therefore, the main parameter that changes with length is $T$. The results of Fig. 6 strongly suggests that one cannot attribute $\Delta$ to an external series resistance, one possible extrinsic parameter that could affect $\Delta$. This is because a series resistance would create a length dependent voltage drop and cause $\Delta$ to decrease with reduced channel length (i.e. with more current), not one that is constant with length as seen in Fig. 6. Although we can't completely rule out the effects from source and drain series resistance, their effects appear to be minimal.

In fact, $\Delta$ is expected to be independent of length if it is an intrinsic parameter of the device. This is because $\Delta$ also represents the carrier concentration near the source, which would be primarily determined by the gate, as we discuss in the Appendix. There, we show that in a quasi-ballistic conductor $\Delta$ and $T$ are one and the same. In a nanoscale MOSFET, the two parameter need to be decoupled since the gate also has an influence on $\Delta$.

### C. *Ballistic Contact Resistance*

We further demonstrate the significance of $T$ with length by extrapolating the purely ballistic resistance value by setting $T=1$. In Fig 7, the black data points represent the channel resistance, $R_{ON}$, of the measured devices. $R_{ON}$ is the output resistance obtained experimentally from the $I_D$-$V_D$ data at a low value of $V_D$ (0.04 V) and at $V_G$=1.2V for $L$< 100nm devices. Extrapolating this linear graph back to the y-axis, or L=0, should give us the contact resistance in the limit of no scattering. This would be proportional to the quantum of resistance (1/Gc) or the contact resistance that arises due to the difference in the number of modes available between the contacts and the channel, modified by the longitudinal field. The contact resistance will be the lowest achievable resistance no matter how advanced the technology for this gate voltage. This extrapolated value is about 170 Ω-μm, or about 85 Ω-μm per contact.

The theoretical minimum value of $R_{ON}$ can also be calculated for each of the four devices by setting the transmission probability to unity ($T$ = 1). These are represented by the red data points in Fig. 7. They are largely independent of length and average to a value of 154 Ω-μm (or 77 Ω-μm per contact), very close to the value obtained experimentally. The similarity in the two values provides important implications on the limit of $R_{ON}$ that can be achieved for our planar devices.

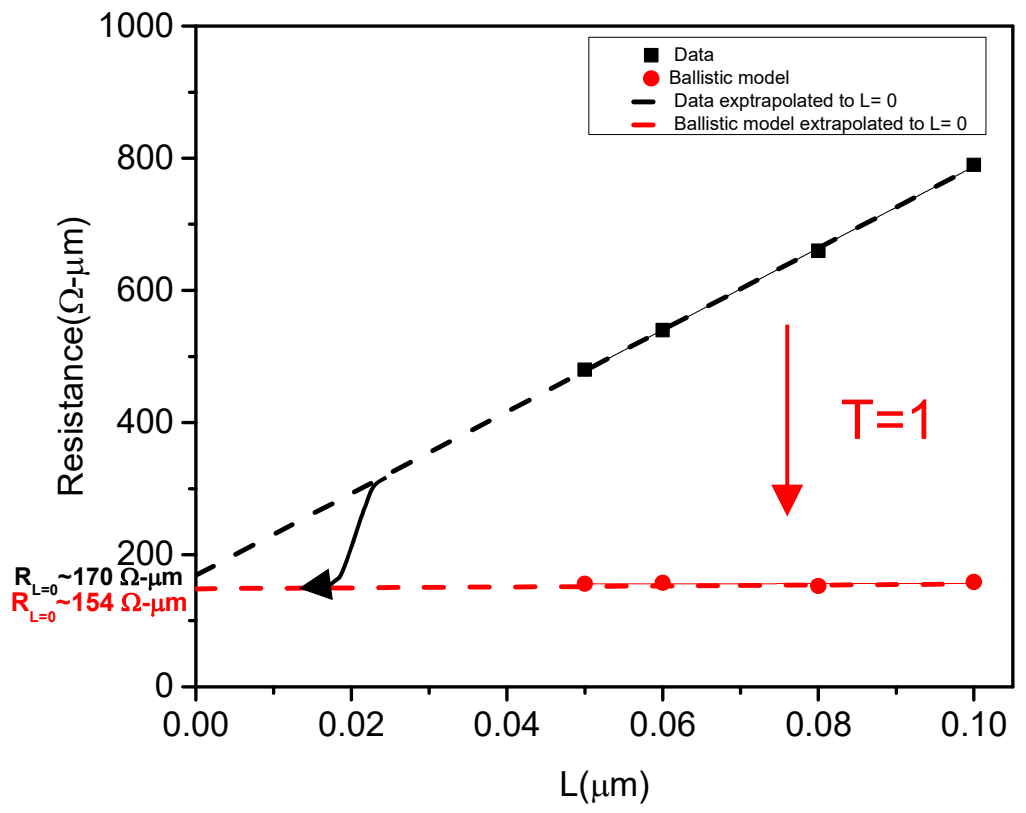

**Fig 7: Experimental channel resistance ($R_{ON}$) measured at $V_D$ =0.04V and $V_G$=1.2V, and purely ballistic $R_{ON}$ by setting T=1 at the same bias. The solid curve is the expected transition to the purely ballistic $R_{ON}$ as L is reduced below λ=19 nm.**

We note that the theoretical value of contact resistance reported in ref. 17 is about half of the value we report. The implications of the larger value we observe are as follows: The inclusion of the drain-channel coupling parameter $\Delta$ has the effect of reducing net carriers that transmit, resulting in a larger theoretical resistance as we show experimentally. $\Delta$ is a function of transistor geometry. We expect $\Delta$ to be smaller in a planar geometry compared to a FinFET geometry. This is because the gate control of the channel is superior in a FinFET [18-21], resulting in a flatter band profile than the one shown in Fig. 2. Therefore, we would expect the theoretical $R_{ON}$ to be smaller for a FinFET resulting from a larger $\Delta$. For the planar geometry, the implication from Fig. 7 is that the transition to the purely ballistic $R_{ON}$ will take place at L ~ λ, as depicted by the solid curve near *L*=19nm.

Finally, we note that including the drain-induced barrier lowering (DIBL) in our model does not qualitatively change the outcome of the results, although the fit to that data improves somewhat. DIBL primarily affects the current in the saturation regime and does little to affect $R_{ON}$, as was noted previously[17]. Including the DIBL parameter adds a small slope to the current in the saturation regime, which is not present in the current model. Our model with DIBL will be the subject of future work.

## V. Conclusion

We report on a two-parameter quasi-ballistic quantum transport model for nanoscale transistors that extends Natori's original ballistic model. The two parameters are *T* and $\Delta$, the transmission probability and drain-channel coupling, respectively. The central feature of our work is the inclusion of $\Delta$, which reduces the net carriers that transmit in the channel and accounts for the spatially varying channel potential. We show that our model can dramatically improve the fit to device data, which we validate by fabricating n-channel MOSFETs of varying lengths. The length scaling of both *T* and $\Delta$ suggests that these are intrinsic parameters of the device, not ones determined by extrinsic effects.

A length dependent analysis extracted from the quasi-ballistic model reveal significant implications regarding the fundamental limit to conductance in a ballistic transistor and the mean free path for carriers in quasi-ballistic MOSFETs. The length scale for the applicability of the model is estimated to be 100 nm and less. We extract a mean scattering length of 19nm at room temperature for our planar devices.

Using the length dependent study, we show that the theoretical $R_{ON}$ is significantly larger than the value one would predict for a device with perfect gate control. The larger $R_{ON}$ arises whenever there is a transverse electric field along the channel, which reduces net modes that transmit compared to a device with no longitudinal electric field in the channel.

## VI. Appendix

Here, we show the origin of the two parameters, *T* and Δ, in our model. First, we illustrate using band diagrams the origin of two quasi Fermi levels along the channel and relate them to quasi-ballistic conductance. We show explicitly that these levels need to be defined locally, as opposed to taking their values from the source/drain contacts.

For a purely ballistic conductor with no scattering, Figure A1 (a) illustrates how the quasi Fermi levels extend from the contacts to the channel (dotted lines). *T*, the transmission probability, is unity, but it can also be viewed as the fraction of the drain quasi-Fermi level that appears in the channel when the source is at ground. Here, the quasi Fermi levels in the channel are those of the source/drain contacts: i.e. the source and drain quasi Fermi levels from the contacts extend into the channel region.

With scattering, $T$ <1. The band diagram that illustrates a uniform transmission along the channel is shown in Fig. A1(b). At the source-end of the channel, $T$ is also the fraction of the drain quasi-Fermi level that appears there.

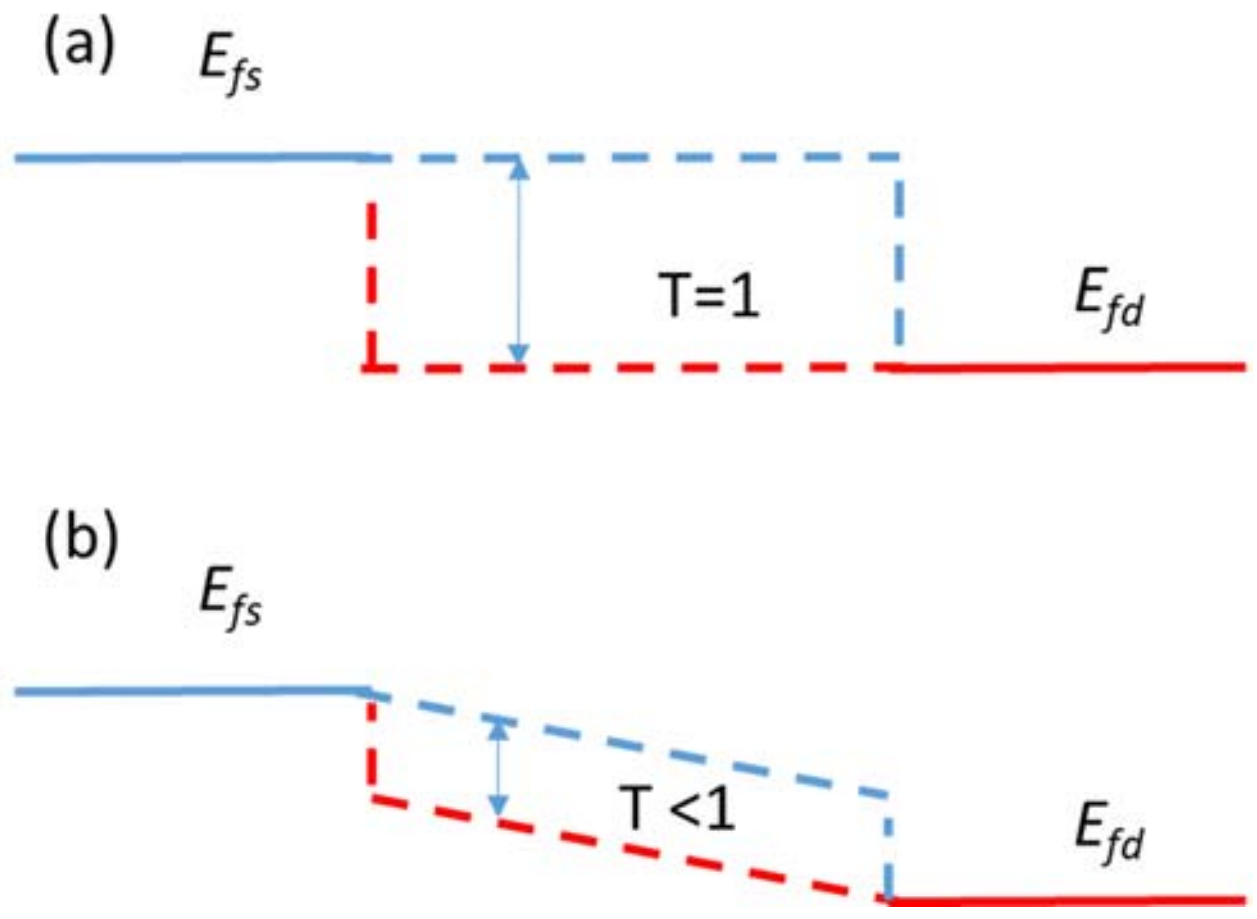

**Fig. A1: Band diagram of (a) a purely ballistic conductor with T=1 and (b) a quasi-ballistic conductor with T<1. The region with dotted quasi Fermi levels represents the channel region of the conductor. Here, T can be viewed as the fraction of $E_{fd}$ that appears on the source-end.**

The band diagrams in Fig. A1 are consistent the derivation of the conductance quantum for a single mode as given in Equations A1 and A2. Eq. A1 can be derived from Eq. (1) in the low temperature limit and by assuming a constant $T$.

$$I = \frac{2e}{h} T\left(E_{fs} - E_{fD}\right) \qquad \text{(A1)}$$

$$G = \frac{2e^2}{h} T \qquad \text{(A2)}$$

The equations above give the correct conductance quantum $G$ as a function of $T$. A similar argument is provided in Ref. 6. Fig. A1(b) would transform to Fig. 1A(b) if $T = 1$. In summary, two quasi-Fermi levels are needed to describe a quasi-ballistic conductor, and one can interpret $T$ as the fraction of the drain quasi-Fermi level that appears at the source-end.

The preceding point of view is different from the typical treatment of diffusive transport where only the local potential in the channel is needed to describe the current flow, as can be seen in the equation for the long-channel FET of Eq. (4). In fact, Fig. A1(a) would also apply to a purely ballistic MOSFET where the gate has perfect control of the channel potential. The same, however, can't be said of Fig. A1(b) for a quasi-ballistic FET where the incomplete coupling of the gate to the channel creates a highly non-uniform potential profile along the channel. Instead a modified band diagram is needed, resulting in Fig. 3 with two parameters to describe the transport as we discuss below.

In a MOSFET, the gate complicates the analysis and one can't use the band diagram for a conductor shown in Fig. A1(b) to represent a quasi-ballistic MOSFET. More importantly, $T$ is no longer the fraction of $E_{fd}$ that appears at the top of the source. If we assume that the gate "pegs" the band profile at the top of the source, then $T$ no longer represents the difference $E_{fs}$-$E_{fd}$ at the peak of the band diagram. Instead, in Fig. 2a, Δ replaces $T$ of Fig. A1(b). Furthermore, Δ is expected to be independent of the channel length for a well-designed MOSFET. Our observation of a constant Δ in Fig. 6 supports this picture.

Thus, we have the origin of the two-parameter fit: T and Δ become decoupled in a quasi-ballistic MOSFET whereas they are one and the same in a quasi-ballistic conductor without a gate.

## VII. Acknowledgements

The authors acknowledge support from DTRA/NRL Grant No. N00173-14-1-G-17 and NSF Grant No. ECCS 1606659. JL would like to thank Prof. M. Lundstrom for useful discussions.

## VIII. Author Contributions

JL conceived the model; RC carried out the fit; RC and PD carried out the measurements; PN fabricated the devices; HH and JL discussed the physics of the model. All authors discussed the data and contributed to writing the manuscript.

*jlee1@sunypoly.edu

## IX. Additional Information

Competing Interests: The authors declare no competing interests.